\documentclass[journal,12pt,onecolumn,draftclsnofoot]{IEEEtran}

\usepackage{color}
\usepackage{multirow}
\usepackage{graphicx}
\usepackage{epstopdf}
\usepackage[cmex10]{amsmath}
\usepackage{amssymb}

\usepackage{multirow}
\usepackage{amsthm}
\usepackage{cite}
\usepackage{acronym} 
\usepackage{algorithm, algcompatible}

\algnewcommand\INPUT{\item[\textbf{Input:}]}
\algnewcommand\OUTPUT{\item[\textbf{Output:}]}

\allowdisplaybreaks

\begin{document}

\title{Extreme Level Crossing Rate: A New Performance Indicator for URLLC Systems} 
\author{Nikolaos~I.~Miridakis,~\IEEEmembership{Senior Member,~IEEE}, Theodoros~A.~Tsiftsis,~\IEEEmembership{Senior Member,~IEEE}, Panagiotis~A.~Karkazis, Helen~C.~Leligou, and Fotis~Foukalas,~\IEEEmembership{Senior Member,~IEEE}
\thanks{\textit{Corresponding Author: T. A. Tsiftsis.}}
\thanks{N.~I.~Miridakis and P.~A.~Karkazis are with the Department of Informatics and Computer Engineering, University of West Attica, Aegaleo 12243, Greece (e-mails: nikozm@uniwa.gr, p.karkazis@uniwa.gr).}
\thanks{T.~A.~Tsiftsis and F.~Foukalas are with the Department of Informatics and Telecommunications, University of Thessaly, 35100 Lamia, Greece (e-mails: \{tsiftsis, foukalas\}@uth.gr).}
\thanks{H.~C.~Leligou is with the Department of Industrial Design and Production Engineering, University of West Attica, Aegaleo 12241, Greece (email: e.leligkou@uniwa.gr).}
}


\maketitle

\begin{abstract}
Level crossing rate (LCR) is a well-known statistical tool that is related to the duration of a random stationary fading process \emph{on average}. In doing so, LCR cannot capture the behavior of \emph{extremely rare} random events. Nonetheless, the latter events play a key role in the performance of ultra-reliable and low-latency communication systems rather than their average (expectation) counterparts. In this paper, for the first time, we extend the notion of LCR to address this issue and sufficiently characterize the statistical behavior of extreme maxima or minima. This new indicator, entitled as extreme LCR (ELCR), is analytically introduced and evaluated by resorting to the extreme value theory and risk assessment. Capitalizing on ELCR, some key performance metrics emerge, i.e., the maximum outage duration, minimum effective duration, maximum packet error rate, and maximum transmission delay. They are all derived in simple closed-form expressions. The theoretical results are cross-compared and verified via extensive simulations whereas some useful engineering insights are manifested.  
\end{abstract}


\IEEEpeerreviewmaketitle

\section{Introduction}
\IEEEPARstart{U}{ltra}-reliable and low-latency communication (URLLC) plays a pivotal role in 5G and beyond-5G wireless networks \cite{j:SaatchiNegin2023,j:MuhammadAlves2022,j:7nines2022,j:OnelHirley2020,j:AlsenwiTran19,j:PopovskiURLLC2019}. It strives to satisfy the stringent reliability and latency requirements of mission- and safety-critical applications, such as communication among machines and robots in Industry 4.0 use cases. Over the last years, both academia and industry aim at the analysis, provision and optimization of the URLLC system performance. The main goal is to achieve reliable communication up to a certain confidence level and, at the same time, to do so at quite a low latency margin. 

Machine and/or reinforcement learning approaches \cite{j:SaatchiNegin2023,j:7nines2022} and classical probabilistic-related methods (e.g., see \cite{j:MuhammadAlves2022,j:OnelHirley2020,j:PopovskiURLLC2019,j:URLLCSurvey2022} and relevant references therein) have been implemented to date. Regarding the latter category, two closely related and well-known performance metrics are the so-called level crossing rate (LCR) and average outage duration (AOD). They both stand as second-order statistics and reflect not only on the communication signal status (quality), but also on the duration or frequency of the said communication status (timeliness). Current state-of-the-art research (e.g., \cite{j:SaatchiNegin2023,j:MuhammadAlves2022,j:YooSeong2020,j:OnelHirley2020}) relies on LCR and AOD to study the behavior of URLLC systems. Nevertheless, LCR and AOD correspond to the system performance \emph{on average} and thereby fail to capture the behavior of \emph{rare} events. Notably, the latter events may drastically influence the communication quality with regards to both reliability and latency in applications with strict demands, such as factory automation and autonomous vehicles.    

Capitalizing on the aforementioned observations, for the first time in this paper, we introduce the notion of extreme LCR (ELCR) to explicitly capture the frequency of signal level crossings with respect to rare (extreme) events. This new performance indicator is directly associated with the maximum or minimum duration a given signal envelope stays below or above a desired threshold value, \emph{up to a certain confidence level; say, a predefined reliability}. To formulate ELCR, we resort to statistical tools from the extreme value theory framework (i.e., distribution convergence of the maximum/minimum case) and risk assessment measures such as the value-at-risk (VaR) and conditional VaR (CVaR). The analysis includes a variety of most popular channel fading conditions and the case of single- as well as multi-antennas at the receiver side. Further, some useful performance metrics are analyzed and evaluated; namely, the minimum effective duration (i.e., minimum duration of successful transmission), maximum outage duration, maximum packet error rate for uncoded transmissions, maximum packet transmission delay, and the associated effective (i.e., mission-critical) transmission rate. Finally, some useful engineering insights manifest the impact of channel fading conditions and the presence of multiple antenna elements on the performance of extremely rare events.\\
 
\noindent \textsl{\textbf{Notation}}:
\begin{itemize}
	\item $\mathbb{E}[\cdot]$ is the expectation operator;
	\item $f_{X}(\cdot)$, $F_{X}(\cdot)$ and $\overline{F_{X}}(\cdot)$ represent the probability density function (PDF), cumulative distribution function (CDF) and complementary CDF of a random variable (RV) $X$, respectively;
	\item $f_{Y|Z}(\cdot|\cdot)$ denotes the PDF of $Y$ conditioned on $Z$ event and $f_{X,Y}(\cdot,\cdot)$ is the joint PDF of $X$ and $Y$ RVs;
	\item $\hat{x}$ denotes an estimate of $x$;
	\item ${\rm li}(\cdot)$ is the logarithmic integral function \cite[Eq. (4.211.2)]{tables};
	\item $\Gamma(\cdot)$ is the Gamma function \cite[Eq. (8.310.1)]{tables} and $\Gamma(\cdot,\cdot)$ is the upper incomplete Gamma function \cite[Eq. (8.350.2)]{tables};
	\item $Q_{n}(\cdot,\cdot)$ is the $n^{\rm th}$-order Marcum-$Q$ function;
	\item $I_{n}(\cdot)$ represents the $n^{\rm th}$-order modified Bessel function of the first kind \cite[Eq. (8.445)]{tables};
	\item $\epsilon$ denotes the Euler-Mascheroni constant \cite[Eq. (8.367)]{tables}.
\end{itemize}

\section{System Model and Channel Statistics}
{\color{black}Consider a wireless communication system with a single-antenna transmitter (e.g., a low-cost sensor device) and multi-antenna receiver (e.g., an access point or base station) equipped with $N\geq 1$ antennas, operating on the uplink direction}. The transmit signal-to-noise ratio (SNR) is defined as $P/N_{0}$ where $P$ and $N_{0}$ denote the transmit power and additive white Gaussian noise power, respectively. Assuming perfect channel-state information (CSI) conditions at the receiver and maximum ratio combining (MRC) detection,\footnote{{\color{black}In the case of downlink transmission, the same system setup may operate via maximum ratio transmission (MRT) to each individual single-antenna device. In such a case, the analytical framework presented in this letter can be directly applied for the downlink transmission counterpart via MRT.}} the received SNR at time instance $t$ reads as 
\begin{align*}
\gamma(t)=\overline{\gamma} \sum^{N}_{n=1}\alpha_{n}^{2}(t),
\end{align*}
where $\gamma(t)$ and $\alpha_{n}(t)$ represent the total received SNR and channel fading envelope of the $n^{\rm th}$ received antenna at time instance $t$, respectively. Also, $\overline{\gamma}\triangleq \frac{P}{N_{0}} d^{-a}$ incorporates the transmit SNR and (deterministic) path losses with $d$ denoting the transceiver distance and $a$ being the path-loss exponent.\footnote{We reasonably assume that $\overline{\gamma}$ is time-independent in the sense that it may not alter after many consecutive timeslots; thereby, it remains known and deterministic with respect to a given time instant $t$.} Let\footnote{Hereinafter, for notational simplicity, we drop symbol $t$ since we refer to any time instance.} 
\begin{align}
\alpha=\left(\overline{\gamma} \sum^{N}_{n=1}\alpha_{n}^{2}\right)^{1/2}\:\: \textrm{and}\:\: \dot{\alpha}=\frac{{\rm d}\alpha}{{\rm d}t}
\end{align}
be the envelope of the total received signal and its time-derivative (i.e., the envelope slope), respectively. Then, for a real stationary process $\alpha$, Rice has shown that \cite{j:RicefamousPaper} 
\begin{align*}
{\rm LCR}_{\alpha}(\alpha_{\rm th})&=\int^{\infty}_{0}x f_{\alpha,\dot{\alpha}}(\alpha_{\rm th},x){{\rm d}x}\\
&=f_{\alpha}(\alpha_{\rm th})\int^{\infty}_{0}x f_{\dot{\alpha}|\alpha}(x|\alpha=\alpha_{\rm th}){{\rm d}x},
\end{align*}
where $\alpha_{\rm th}$ denotes a given envelope threshold value and ${\rm LCR}_{\alpha}(\cdot)$ is the LCR of $\alpha$ (i.e., the average number of times where $\alpha$ crosses $\alpha_{\rm th}$ in the positive direction per second). 

For the most popular small-scale channel fading models (such as Rayleigh, Rician and Nakagami), independence between $\alpha$ and $\dot{\alpha}$ under certain conditions holds such that
\begin{align}
f_{\alpha,\dot{\alpha}}(\alpha_{\rm th},x)=f_{\alpha}(\alpha_{\rm th})f_{\dot{\alpha}}(x).
\label{independentJointPDF}
\end{align} 
The sufficient conditions of \eqref{independentJointPDF} dictate that (\textit{a}.) the phase distribution of scatter components of $\alpha$ should be symmetric about $\pi$ (e.g., isotropic scattering is a particular example) and (\textit{b}.) line-of-sight (LoS) component, whenever exists, should be predetermined (moving yet known) or fixed \cite{b:PrinciplesofMobileCommunications2017,j:BeaulieuDong2003}. The distribution functions of $\alpha$, {\color{black}for independent and identically distributed faded channels}, are expressed as \cite{b:DigCom2005}
\begin{align}
f_{\alpha}(x)=\left\{\begin{array}{c l}     
    \frac{2 x^{N}\left(\frac{K+1}{\overline{\gamma}}\right)^{\frac{N+1}{2}}I_{N-1}\left(2 \sqrt{\frac{K N (K+1) x^{2}}{\overline{\gamma}}}\right)}{(K N)^{\frac{N-1}{2}}\exp\left(K N+\frac{(K+1) x^{2}}{\overline{\gamma}}\right)},& \textrm{Rician},\\
		\\
		\frac{2 x^{2 N m-1}}{\Gamma(N m) (\overline{\gamma}/m)^{N m}}\exp\left(-\frac{m x^{2}}{\overline{\gamma}}\right),& \textrm{Nakagami},
\end{array}\right.
\label{aPDF}
\end{align} 
and
\begin{align}
F_{\alpha}(x)=\left\{\begin{array}{c l}     
    1-Q_{N}\left(\sqrt{2 K N},\sqrt{\frac{2 (K+1) x^{2}}{\overline{\gamma}}}\right),& \textrm{Rician},
		\\
    1-\frac{\Gamma\left(N m,\frac{m x^{2}}{\overline{\gamma}}\right)}{\Gamma(N m)},& \textrm{Nakagami},
\end{array}\right.
\label{aCDF}
\end{align}
where $K\geq 0$ stands for the Rician$-K$ factor which determines the power ratio of the specular component over the random scatterers. Also, $m\geq 1/2$ is the Nakagami$-m$ parameter. For $N=1$, \eqref{aPDF} and \eqref{aCDF} reduce to the scenario of single-antenna transceiver communication.

For the aforementioned sufficient conditions of \eqref{independentJointPDF}, the envelope time-derivative follows a zero-mean Gaussian distribution with variance $\dot{\sigma}^{2}$ yielding \cite{b:DigCom2005}
\begin{align}
f_{\dot{\alpha}}(x)=\frac{\exp(-\frac{x^{2}}{2 \dot{\sigma}^{2}})}{\sqrt{2 \pi} \dot{\sigma}}\:\: \textrm{with}\:\: \dot{\sigma}^{2}=\frac{\overline{\gamma}}{\rm B} \pi^{2} f^{2}_{\rm m},
\label{sdot}
\end{align}
where $f_{\rm m}$ is the maximum Doppler frequency shift and ${\rm B}\triangleq K+1$ for Rician fading or ${\rm B}\triangleq m$ for Nakagami fading. By setting $K=0$ (or $m=1$) at ${\rm B}$ in \eqref{sdot}, the variance of the envelope time-derivative for the Rayleigh channel fading case is obtained. Then, LCR becomes ${\rm LCR}_{\alpha}(\alpha_{\rm th})=f_{\alpha}(\alpha_{\rm th})\int^{\infty}_{0}x f_{\dot{\alpha}}(x){{\rm d}x}=f_{\alpha}(\alpha_{\rm th})\dot{\sigma}/\sqrt{2 \pi}$.
With LCR given at hand, another powerful statistical tool is the so-called AOD (say, the average time that envelope $\alpha$ stays below $\alpha_{\rm th}$), which is defined as the reciprocal of LCR conditioned on its CDF as ${\rm AOD}_{\alpha}(\alpha_{\rm th})=F_{\alpha}(\alpha_{\rm th})/{\rm LCR}_{\alpha}(\alpha_{\rm th})$ (in seconds). In a similar basis, the average time that envelope $\alpha$ stays above $\alpha_{\rm th}$ entitled as average effective duration (AED) reads as ${\rm AED}_{\alpha}(\alpha_{\rm th})=\overline{F_{\alpha}}(\alpha_{\rm th})/{\rm LCR}_{\alpha}(\alpha_{\rm th})$.

\section{Statistics of Extreme Channel Conditions}
Although LCR and AOD are useful second-order channel statistics, they cannot sufficiently capture the behavior of rare events (occurring at the distribution tail) since--by definition--they reflect only on the \emph{expected value} of the relevant performance metric. Paradigms of paramount interest are the maximum outage duration and the minimum duration of successful transmission. Unfortunately, these metrics cannot be described via the conventional LCR and AOD statistics. To this end, we resort to the extreme value theory \cite{b:Haan2006} which efficiently describes rare events. Accordingly, the maximum or minimum of asymptotically many independent and identically distributed random events belongs to the maximum domain of attraction (MDA) of one of the three possible distribution types; namely the Gumbel, Fr\'echet or Weibull. Insightfully, Gaussian distribution belongs to the MDA of Gumbel distribution and is uniquely recovered therein \cite[Thm.~5]{j:SHIMURA2012}.\footnote{The explicit definition of a recoverable distribution is provided in \cite[Eq. (1.1)]{j:SHIMURA2012}.} Hence, we define the (asymptotic) minimum and maximum $\dot{\alpha}$, respectively, as 
\begin{align*}
\dot{\alpha}_{\min}\triangleq \min_{i}\{\dot{\alpha}_{i}\}^{\infty}_{i=1}\:\:\textrm{and}\:\: \dot{\alpha}_{\max}\triangleq \max_{i}\{\dot{\alpha}_{i}\}^{\infty}_{i=1}.
\end{align*}
Then, according to \eqref{sdot}, $f_{\dot{\alpha}_{\min}}(\cdot)$ and $f_{\dot{\alpha}_{\max}}(\cdot)$ both converge to the Gumbel distribution in the limit; yet with slightly different parametric values, as it will subsequently be shown.

The Gumbel CDF of the maximum event is defined as \cite{j:Gumbel1960a}
\begin{align}
F_{\rm G}(x;\mu_{\rm G,\max},\sigma_{\rm G,\max})=\exp\left(-\exp\left(-\frac{x-\mu_{\rm G,\max}}{\sigma_{\rm G,\max}}\right)\right),
\label{cdfGumbel}
\end{align}
where $\mu_{\rm G,\max}$ and $\sigma_{\rm G,\max}$ are the so-called location and scale parameter, respectively. Also, the expected value and variance of Gumbel distribution are given as $\mu_{\rm G,\max}-\epsilon \sigma_{\rm G,\max}$ and $\sigma^{2}_{\rm G,\max} \pi^{2}/6$, correspondingly. {\color{black}According to \cite[Lemma~2]{j:SongLi2006}, convergence in (asymptotic) distribution results to moment convergence; i.e., $\mu_{\rm G,\max}-\epsilon \sigma_{\rm G,\max}=\mathbb{E}[\dot{\alpha}]$ and $\sigma^{2}_{\rm G,\max} \pi^{2}/6=\mathbb{E}[\dot{\alpha}^{2}]-(\mathbb{E}[\dot{\alpha}])^{2}$. Via the method of moments-matching estimation, while noticing from \eqref{sdot} that $\mathbb{E}[\dot{\alpha}]=0$ and $\mathbb{E}[\dot{\alpha}^{2}]=\dot{\sigma}^{2}$}, we obtain their respective estimates as
\begin{align}
\hat{\mu}_{\rm G,\max}=\epsilon \frac{\sqrt{6 \dot{\sigma}^{2}}}{\pi}\:\textrm {and}\: \hat{\sigma}_{\rm G,\max}=\frac{\sqrt{6 \dot{\sigma}^{2}}}{\pi}.
\label{mGsG}
\end{align}
Thus, the distribution of $\dot{\alpha}_{\max}$ is approached by
\begin{align}
F_{\dot{\alpha}_{\max}}(x)=F_{\rm G}(x;\hat{\mu}_{\rm G,\max},\hat{\sigma}_{\rm G,\max}).
\label{cdfMax}
\end{align}

Provided with the tail statistics of envelope time-derivative, $\dot{\alpha}_{\max}$ can be further quantified given a specific \emph{level of risk}. By introducing the notion of confidence level $q$, we are able to characterize the level of risk in terms of the number of envelope crossings per unit time; for instance, a confidence level of $q$ means that there is a $q\%$ possibility that the worst-case $\dot{\alpha}$ will not exceed a certain time threshold. Typical percentile values of the confidence level are $q=99\%$ or $q=95\%$. Nonetheless, it may reach up to $q=99.99999\%$ ($7-$nines) for some ultra-reliable applications \cite{j:7nines2022}. VaR is regarded as a key performance metric to quantify the effect of rare events. It is defined as \cite[Def.~3.3]{j:Artzner1999}
\begin{align}
\nonumber
{\rm VaR}_{\dot{\alpha}_{\max}}(q)&\triangleq \mathcal{Q}_{\dot{\alpha}_{\max}}(q)\\
&=\hat{\mu}_{\rm G,\max}-\hat{\sigma}_{\rm G,\max}{\rm ln}\left(-{\rm ln}(q)\right),\quad 0<q<1,
\label{VAR}
\end{align}
where $\mathcal{Q}_{\dot{\alpha}_{\max}}(\cdot)$ is the quantile function of $\dot{\alpha}_{\max}$. However, VaR is an incoherent risk measure.\footnote{A coherent risk metric satisfies certain properties regarding its supporting function; namely, translational invariance, sub-additivity, monotonicity, and homogeneity \cite{j:Artzner1999}.} An alternative measure, which is indeed coherent and more reliable than VaR is the so-called CVaR. It is presented as \cite[Prop.~15]{j:Norton2021}
\begin{align}
\nonumber
{\rm CVaR}_{\dot{\alpha}_{\max}}(q)&\triangleq \frac{1}{1-q}\int^{1}_{q}{\rm VaR}_{\dot{\alpha}_{\max}}(y)dy\\
\nonumber
&=\hat{\mu}_{\rm G,\max}+\frac{\hat{\sigma}_{\rm G,\max}}{1-q}\big[{q {\rm ln}\left(-{\rm ln}(q)\right)-\rm li}(q)+\epsilon \big]\\
&=f_{\rm m} \sqrt{\frac{6 \overline{\gamma}}{\rm B}}\left(\frac{\Xi}{1-q}+\epsilon\right),
\label{CVAR}
\end{align}
where $\Xi\triangleq {q {\rm ln}\left(-{\rm ln}(q)\right)-\rm li}(q)+\epsilon$ is introduced for notational clarity. The proof of \eqref{CVAR} is relegated in Appendix~\ref{appa}. It is noteworthy that ${\rm CVaR}_{\dot{\alpha}_{\max}}(q)$ is the expectation on the worst $(1-q)\%$ values of $\dot{\alpha}_{\max}$, which reflects that it is more sensitive than VaR to the shape of $\dot{\alpha}$ distribution in the right tail. 

For the other extreme case of the minimum event $\dot{\alpha}_{\min}$ (in the left tail), the corresponding distribution function is given in \eqref{cdfGumbel} by simply replacing $x$ with $-x$. By doing so, we arrive at the following statistics:
\begin{align}
\hat{\mu}_{\rm G,\min}=-\epsilon \frac{\sqrt{6 \dot{\sigma}^{2}}}{\pi}\:\textrm{and}\: \hat{\sigma}_{\rm G,\min}=\hat{\sigma}_{\rm G,\max}=\frac{\sqrt{6 \dot{\sigma}^{2}}}{\pi},
\label{mG2sG2}
\end{align}
\begin{align}
{\rm VaR}_{\dot{\alpha}_{\min}}(q)=\hat{\mu}_{\rm G,\min}-\hat{\sigma}_{\rm G,\min}{\rm ln}\left(-{\rm ln}(1-q)\right),
\label{VAR2}
\end{align}
and
\begin{align}
\nonumber
&{\rm CVaR}_{\dot{\alpha}_{\min}}(q)\triangleq \frac{1}{q}\int^{q}_{0}{\rm VaR}_{\dot{\alpha}_{\min}}(1-y)dy\\
\nonumber
&=\hat{\mu}_{\rm G,\min}+\frac{\hat{\sigma}_{\rm G,\min}}{q}\big[{(1-q) {\rm ln}\left(-{\rm ln}(1-q)\right)-\rm li}(1-q)+\epsilon\big]\\
&=f_{\rm m} \sqrt{\frac{6 \overline{\gamma}}{\rm B}}\left(\frac{\Psi}{q}-\epsilon\right),
\label{CVAR2}
\end{align}
where $\Psi\triangleq {(1-q) {\rm ln}\left(-{\rm ln}(1-q)\right)-\rm li}(1-q)+\epsilon$. Notably, the terms $\Xi$ and $\Psi$ within \eqref{CVAR} and \eqref{CVAR2} can be a priori computed and therefore are regarded as an \emph{offline} operation.

Meanwhile, by leveraging on the independence between the received envelope and its extreme time-derivative (maximum or minimum envelope slope with respect to time), we introduce the extreme LCRs (in crossings per second) for a confidence level of $q$. In analogy to the LCR definition, they expressed as
\begin{align}
{\rm ELCR}_{\dot{\alpha}_{\max}}(\alpha_{\rm th};q)\triangleq f_{\alpha}(\alpha_{\rm th}){\rm CVaR}_{\dot{\alpha}_{\max}}(q),
\label{ELCRmax}
\end{align}   
and
\begin{align}
{\rm ELCR}_{\dot{\alpha}_{\min}}(\alpha_{\rm th};q)\triangleq f_{\alpha}(\alpha_{\rm th}){\rm CVaR}_{\dot{\alpha}_{\min}}(q).
\label{ELCRmin}
\end{align}  
More details on the above ELCR derivations are provided in Appendix~\ref{appb}. Notably, \eqref{ELCRmax} or \eqref{ELCRmin} express the maximum or minimum number of crossings per second, respectively, at a given threshold $\alpha_{\rm th}$ for a certain confidence level $q$.

Based on the latter ELCR expressions, some useful performance indicators arise; namely the maximum outage duration ${\rm OD}_{\max}$ and the minimum effective duration ${\rm ED}_{\min}$ (i.e., minimum duration of successful transmission). They are respectively presented (and both measured in seconds) as
\begin{align}
{\rm OD}_{\max}(\alpha_{\rm th};q)\triangleq \frac{F_{\alpha}(\alpha_{\rm th})}{f_{\alpha}(\alpha_{\rm th}){\rm CVaR}_{\dot{\alpha}_{\min}}(q)},
\label{MOD}
\end{align}
and
\begin{align}
{\rm ED}_{\min}(\alpha_{\rm th};q)\triangleq \frac{\overline{F_{\alpha}}(\alpha_{\rm th})}{f_{\alpha}(\alpha_{\rm th}){\rm CVaR}_{\dot{\alpha}_{\max}}(q)}.
\label{ED}
\end{align}

\section{Performance Metrics}
Capitalizing on the derivation of ELCR, we subsequently focus on key performance metrics; namely the maximum packet error rate (MPER), maximum packet transmission delay (MPTD) and effective throughput. We commence by analyzing MPER first. Uncoded transmissions are assumed (viz., a packet error is declared if at least one bit error is detected in the packet) and a fixed packet transmission time $\rm T_{\rm p}$ (typically, $\rm T_{\rm p}$ is in the order of $10^{-3}$ seconds for low-latency services). 

Error rate (and other performance metrics, e.g., outage and system throughput) is usually associated with the received SNR rather than its corresponding envelope. To this end, recall that $\gamma=\alpha^{2}{\color{black}(=\overline{\gamma} \sum^{N}_{n=1}\alpha_{n}^{2})}$ and thereby $\alpha_{\rm th}=\sqrt{\gamma_{\rm th}}$. According to \cite[Eq. (59)]{j:Fukawa12}, the packet error rate for a given SNR target $\gamma_{\rm th}$ is directly related to the traditional LCR such that ${\rm PER}(\gamma_{\rm th})=F_{\alpha}(\sqrt{\gamma_{\rm th}})\exp(-{\rm LCR}_{\alpha}(\sqrt{\gamma_{\rm th}}) {\rm T_{\rm p}}/\overline{F_{\alpha}}(\sqrt{\gamma_{\rm th}}))$. In a similar basis, we define the MPER with a $q-$confidence level by invoking \eqref{ELCRmax} which is given by
\begin{align}
\nonumber
&{\rm MPER}(\gamma_{\rm th};q)=F_{\alpha}(\sqrt{\gamma_{\rm th}}) \exp\left(-\frac{{\rm ELCR}_{\dot{\alpha}_{\max}}(\sqrt{\gamma_{\rm th}};q) {\rm T_{\rm p}}}{\overline{F_{\alpha}}(\sqrt{\gamma_{\rm th}})}\right)\\
&=F_{\alpha}(\sqrt{\gamma_{\rm th}})\exp\left(-\frac{{\rm T_{\rm p}} f_{\rm m} \sqrt{\frac{6 \overline{\gamma}}{\rm B}}\left(\frac{\Xi}{1-q}+\epsilon\right)f_{\alpha}(\sqrt{\gamma_{\rm th}})}{\overline{F_{\alpha}}(\sqrt{\gamma_{\rm th}})}\right).
\label{MPER}
\end{align}
MPER in \eqref{MPER} is interpreted as the worst-case PER (reflected by the most rapid envelope changes, i.e., maximum envelope slope $\dot{\alpha}$) within time period ${\rm T_{\rm p}}$ at a $q\%$ reliability.

Generally in URLLC systems, we aim to satisfy ${\rm MPER}(\gamma_{\rm th};q)\leq 1-q$. The case when the signal undergoes Rayleigh channel fading and $N=1$ (i.e., single-antenna receiver) admits a closed-form expression with respect to the SNR threshold. Letting $\gamma_{\rm th}\triangleq 2^{\mathcal{R}}-1$, where $\mathcal{R}$ is the target transmission rate (in bps/Hz), we get after some straightforward manipulations 
\begin{align}
\nonumber
&\frac{\sqrt{\gamma_{\rm th}}}{\overline{\gamma}}\left[\sqrt{\gamma_{\rm th}}+2 {\rm T}_{\rm p} f_{\rm m} \sqrt{\frac{6 \overline{\gamma}}{\rm B}}\left(\frac{\Xi}{1-q}+\epsilon\right) \right]\leq -{\rm ln}(q) \Leftrightarrow\\
\nonumber
&\mathcal{R}\leq {\rm log}_{2}\Bigg(1+\Bigg[\sqrt{-{\rm ln}(q)\overline{\gamma}+\left({\rm T}_{\rm p}f_{\rm m} \sqrt{\frac{6 \overline{\gamma}}{\rm B}}\left(\frac{\Xi}{1-q}+\epsilon\right)\right)^{2}}\\
&-{\rm T}_{\rm p} f_{\rm m} \sqrt{\frac{6 \overline{\gamma}}{\rm B}}\left(\frac{\Xi}{1-q}+\epsilon\right)\Bigg]^{2}\Bigg).
\label{MPTDRayN1}
\end{align}
On the other hand, if the goal is the throughput maximization, the suitable $\mathcal{R}$ is directly computed by turning the inequality of \eqref{MPTDRayN1} into equality; entitled as $\mathcal{R}^{\star}$. Notably, the right-hand side of \eqref{MPTDRayN1} serves as a lower bound of $\mathcal{R}^{\star}$ whenever a LoS signal is present and/or $N>1$. Regarding the general setup of $N$ receive antennas and/or more versatile channel fading models (such as Rician or Nakagami), $\mathcal{R}^{\star}$ can be numerically solved utilizing \eqref{MPER} as
\begin{align}
\textstyle \overline{F_{\alpha}}(\sqrt{2^{\mathcal{R}^{\star}}-1})\exp\left(-\frac{{\rm T_{\rm p}} f_{\rm m} \sqrt{\frac{6 \overline{\gamma}}{\rm B}}\left(\frac{\Xi}{1-q}+\epsilon\right)f_{\alpha}(\sqrt{2^{\mathcal{R}^{\star}}-1})}{\overline{F_{\alpha}}(\sqrt{2^{\mathcal{R}^{\star}}-1})}\right)=q.
\label{Ropt}
\end{align}
It is noteworthy that $\mathcal{R}^{\star}$ is the maximum achievable transmission rate so as to satisfy $q\%$ reliability; say, $\mathcal{R}^{\star}(q)$ which is a compact and suitable solution for URLLC systems. 

Additionally, the transmission delay denotes a critical performance indicator to preserve the system design efficiency. The maximum packet transmission delay is defined as the time required for a successful packet transmission with $q\%$ reliability. If $l$ transmissions occur, this means that $(l-1)$ consecutive transmissions were unsuccessful, while the $l^{\rm th}$ transmission was successful. It is mathematically expressed (unconditioned on $l$) as
\begin{align}
\nonumber
&{\rm MPTD}(\gamma_{\rm th};q)\\
\nonumber
&={\rm T}_{\rm p}\sum^{\infty}_{l=1}l \times {\rm MPER}(\gamma_{\rm th};q)^{l-1}[1-{\rm MPER}(\gamma_{\rm th};q)]\\
&={\rm T}_{\rm p}[1-{\rm MPER}(\gamma_{\rm th};q)]^{-1}.
\end{align}
Putting \eqref{MPER} into the latter expression, MPTD is obtained in a straightforward closed-form expression. Both MPER and MPTD can be applied for most small-scale channel fading distributions of practical interest by correspondingly plugging \eqref{aPDF} and \eqref{aCDF}.

\section{Numerical Results}
In this section, the derived analytical results are verified via numerical validation (in line-curves), whereas they are cross-compared with corresponding Monte-Carlo simulations (in solid-circle marks). Without loss of generality and for the sake of clarity, hereinafter, we focus on the Rician channel fading case. We set the maximum Doppler frequency $f_{\rm m}=100$Hz and the confidence level $q=99.999\%$ (unless otherwise specified). The method proposed in \cite{rockafellar2002conditional} is used for the computation of CVaR (which is required for the ELCR derivations).

\begin{figure}[!t]
\centering
\includegraphics[trim=2.0cm .1cm 3.0cm .2cm, clip=true,totalheight=0.4\textheight]{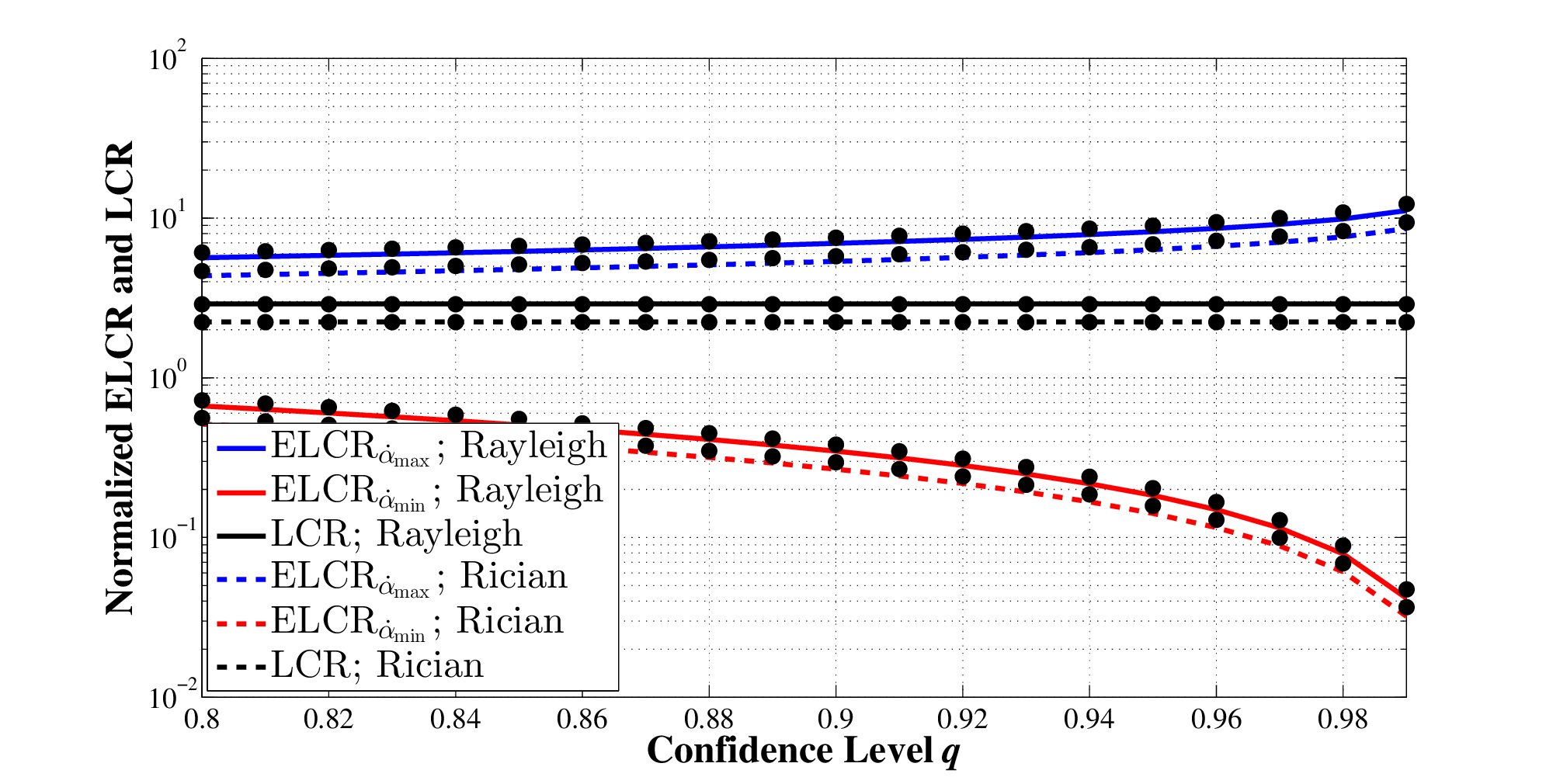}
\caption{Normalized (i.e, $\times 1/f_{\rm m}$) ELCRs and conventional LCR vs. various confidence levels. For the Rician fading case, $K=10$dB is assumed. Also, $\gamma_{\rm th}=\overline{\gamma}=10$dB and $N=1$ (i.e., a single-antenna transceiver).}
\label{fig1}
\end{figure}

\begin{figure}[!t]
\centering
\includegraphics[trim=2.0cm .1cm 3.0cm .2cm, clip=true,totalheight=0.4\textheight]{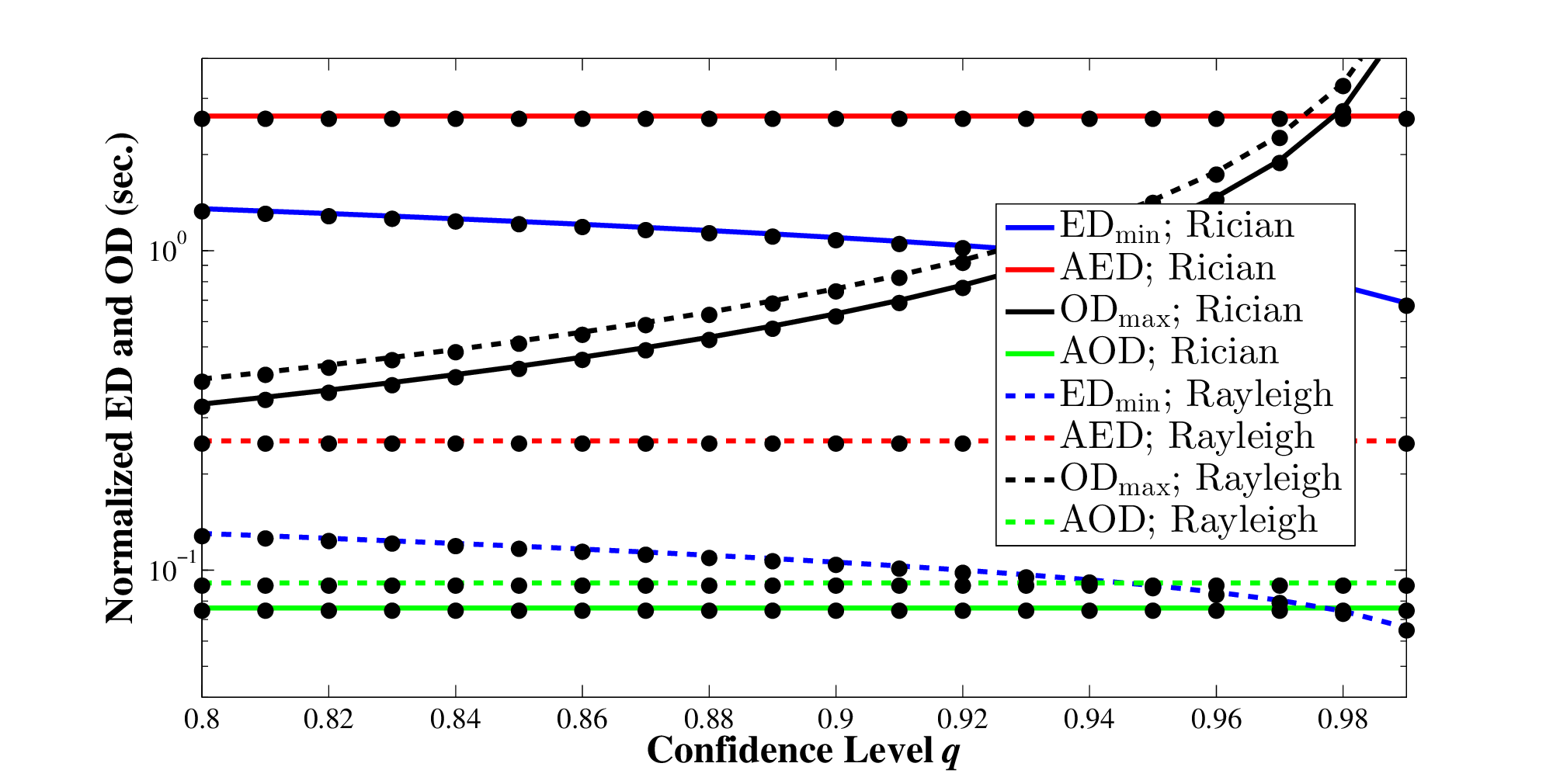}
\caption{{\color{black}Normalized (i.e, $\times f_{\rm m}$) maximum outage duration ${\rm OD}_{\max}$, minimum effective duration ${\rm ED}_{\min}$, AOD, and AED vs. various confidence levels. For the Rician fading case, $K=10$dB is assumed. Also, $\gamma_{\rm th}=\overline{\gamma}=10$dB and $N=2$.}}
\label{newfig2}
\end{figure}

In Fig.~\ref{fig1}, the difference of the extreme level crossing rates of rare events in comparison to the conventional (average) LCR is evident. It is worthy to state that outliers around the average value (say, LCR) impact more emphatically at a higher reliability percentage since ${\rm ELCR}_{\dot{\alpha}_{\max}}$ (${\rm ELCR}_{\dot{\alpha}_{\min}}$) defines the expectation on the $(1-q)\%$ highest (lowest) scores on level crossings per second. Moreover, the presence of a strong LoS component (when $K=10$) results to a more `concentrated' received channel gain (i.e., fluctuation occurs around a non-zero and static mean) yielding less crossings than in a Rayleigh channel (when $K=0$). {\color{black}These outcomes are further verified in Fig.~\ref{newfig2}, where the extreme outage and effective durations are illustrated and compared with the corresponding conventional average counterparts.}  
 
\begin{figure}[!t]
\centering
\includegraphics[trim=2.0cm .1cm 3.0cm .2cm, clip=true,totalheight=0.4\textheight]{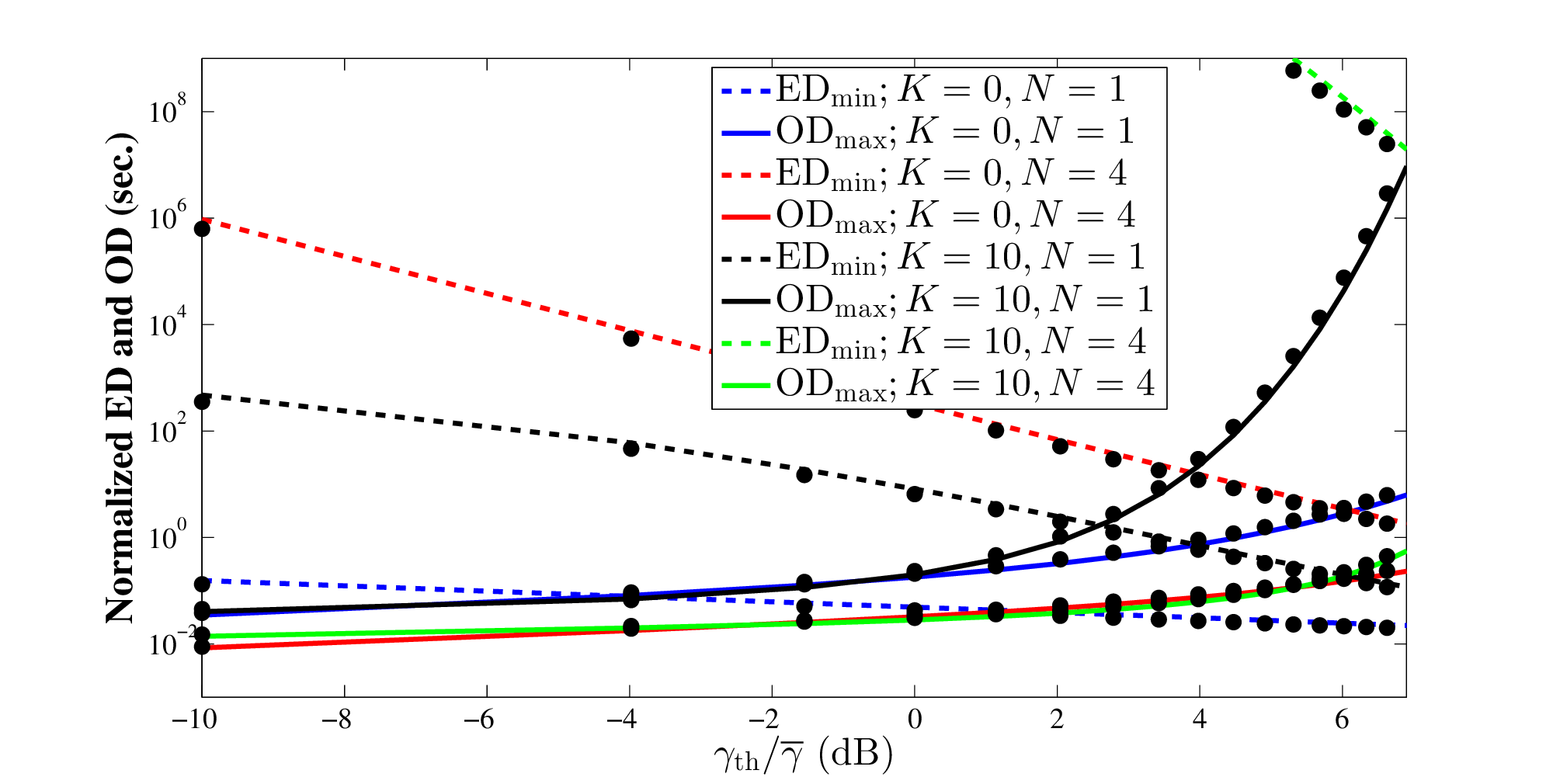}
\caption{Normalized (i.e, $\times f_{\rm m}$) maximum outage duration ${\rm OD}_{\max}$ and minimum effective duration ${\rm ED}_{\min}$  vs. various normalized SNR threshold values for different system setups.}
\label{fig2}
\end{figure}

\begin{figure}[!t]
\centering
\includegraphics[trim=2.0cm .1cm 3.0cm .2cm, clip=true,totalheight=0.4\textheight]{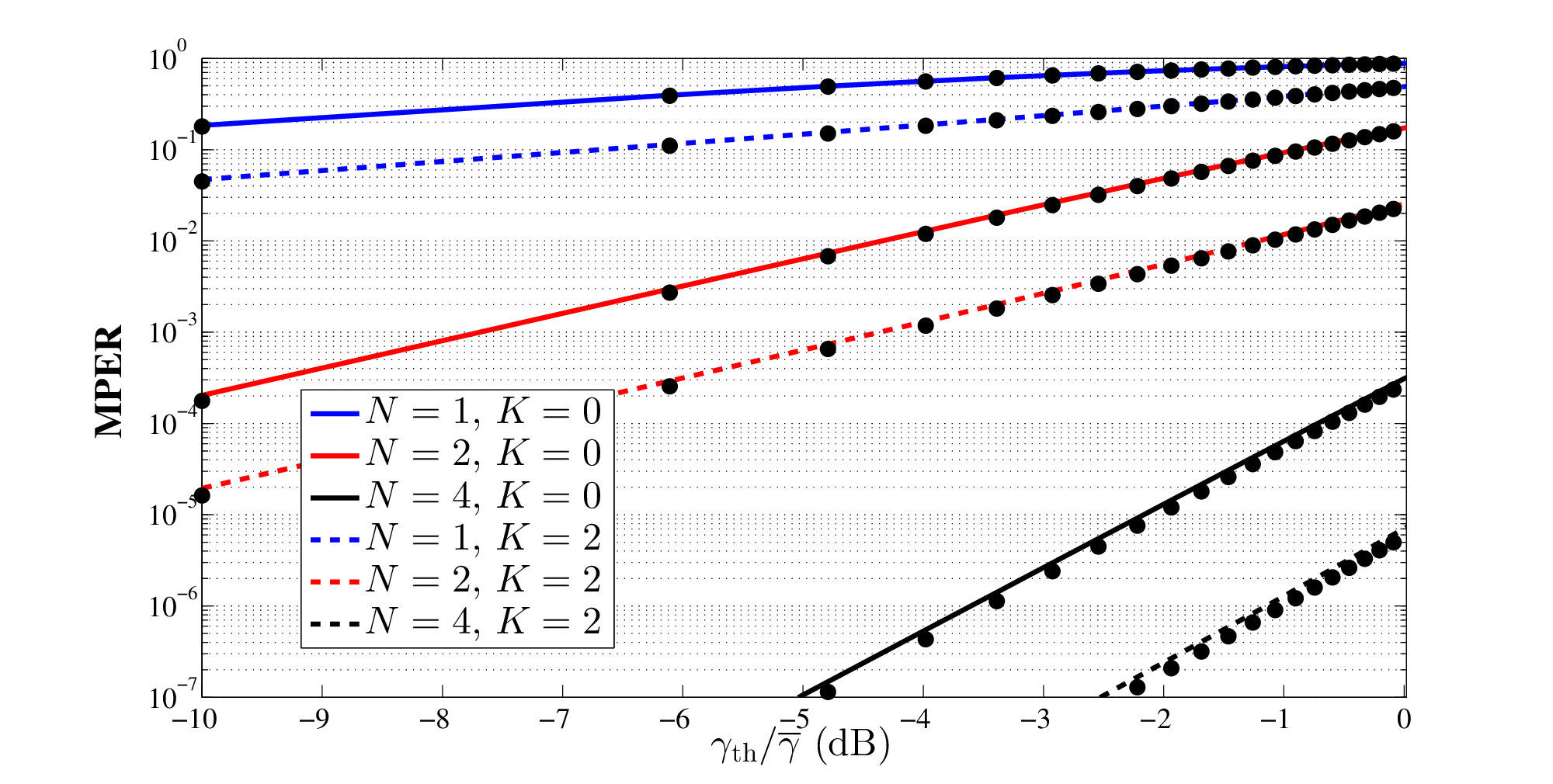}
\caption{MPER  vs. various normalized SNR threshold values for different system setups and ${\rm T}_{\rm p}=1$ms.}
\label{fig3}
\end{figure}

In Fig.~\ref{fig2}, the maximum outage duration ${\rm OD}_{\max}$ and minimum effective duration ${\rm ED}_{\min}$ are illustrated for various normalized SNR threshold values. As expected, the former metric is decreased while the latter one is increased for a higher SNR threshold. Both metrics are being enhanced by the presence of a strong LoS signal and/or a multi-antenna array at the receiver. Interestingly, a higher antenna volume (i.e., higher $N$) seems to be more impactful than the presence of LoS signal propagation (i.e., higher Rician$-K$ factor). In turn, this corresponds to an overall system performance improvement thanks to the underlying spatial diversity provided. The latter outcomes are further verified in Fig.~\ref{fig3} where the maximum packet error rate MPER is illustrated for different system setups. Obviously, the performance gain obtained from an increased antenna array is pronounced.

{\color{black}\section{Conclusion}
The novel concept of ELCR was introduced, analyzed and evaluated in this paper. ELCR is able to capture the performance of extremely rare events and how they reflect on the duration of a faded signal below or above a predetermined threshold value at any intended confidence level. Thereby, it may find direct application in delay-critical and/or high reliable services, such as the URLLC systems. The derived results can be applied on most popular channel fading models (i.e., Rician and Nakagami types of fading) as well as an arbitrary range of antenna elements at the receiver side. Insightfully, the antenna array volume seem to play a more critical role in the enhancement of system performance (by compensating extreme channel fading outliers) than the underlying channel fading type itself.}

\appendix
\subsection{Derivation of \eqref{CVAR} and \eqref{CVAR2}}
\label{appa}
\numberwithin{equation}{subsection}
\setcounter{equation}{0}
In \eqref{CVAR}, an integral of the following form appears: 
\begin{align}
\mathcal{I}\triangleq \int^{1}_{q}{\rm ln}(-{\rm ln}(y))dy=\int^{-{\rm ln}(q)}_{0}{\rm ln}(u)\exp(-u)du,
\end{align}
where the last equality is obtained by implementing integration by substitution and setting $u=-{\rm ln}(y)$. By utilizing \cite[Eq. (1.6.10.1)]{PrudnikovVol1}, we reach at $\int {\rm ln}(u)\exp(-u)du=-\epsilon-e^{-u}{\rm ln}(u)-\Gamma(0,u)$. Then, $\mathcal{I}$ is evaluated at the points of interest as
\begin{align}
\nonumber
\mathcal{I}&=-\epsilon-\frac{{\rm ln}(u)}{e^{u}}-\Gamma(0,u)|^{-{\rm ln}(q)}_{0}\\
&=-\epsilon-q {\rm ln}(-{\rm ln}(q))+{\rm li}(q),
\end{align}
where the last transformation arises by utilizing \cite[Eq. (8.359.2)]{tables}. Thus, \eqref{CVAR} follows after straightforward simplifications. Following similar steps as in solving $\mathcal{I}$ above, CVaR of the other extreme case in \eqref{CVAR2} is derived. 

\subsection{Formulation of the ELCR expressions}
\label{appb}
\numberwithin{equation}{subsection}
\setcounter{equation}{0}
The expected amount of time the envelope lies within the interval $\{\alpha_{\rm th},\alpha_{\rm th}+{\rm d}\alpha\}$ for a (given) maximum envelope slope $\dot{\alpha}_{\max}$ and time increment ${\rm d}t$ is $f_{\alpha,\dot{\alpha}_{\max}}(\alpha_{\rm th},\dot{\alpha}_{\max}){\rm d}\alpha\: {\rm d}\dot{\alpha}_{\max}\: {\rm d}t$. Also, the time required for the envelope to traverse the interval $\{\alpha_{\rm th},\alpha_{\rm th}+{\rm d}\alpha\}$ once for the given slope $\dot{\alpha}_{\max}$ is ${\rm d}\alpha/\dot{\alpha}_{\max}$. The ratio of these quantities is the expected number of envelope crossings in the interval $\{\alpha_{\rm th},\alpha_{\rm th}+{\rm d}\alpha\}$ with regards to a given maximum slope $\dot{\alpha}_{\max}$ and time increment ${\rm d}t$. For a unit time interval, this ratio becomes $\dot{\alpha}_{\max} f_{\alpha,\dot{\alpha}_{\max}}(\alpha_{\rm th},\dot{\alpha}_{\max}){\rm d}\dot{\alpha}_{\max}=f_{\alpha}(\alpha_{\rm th})\dot{\alpha}_{\max} f_{\dot{\alpha}_{\max}}(\dot{\alpha}_{\max}){\rm d}\dot{\alpha}_{\max}$. Hence, the conditional expectation of the number of crossings regarding the $(1-q)\%$ highest $\dot{\alpha}_{\max}$ values (with a positive slope) reads as 
\begin{align}
\nonumber
{\rm ELCR}_{\dot{\alpha}_{\max}}(\alpha_{\rm th};q)&=\frac{f_{\alpha}(\alpha_{\rm th})}{1-q}\int^{\infty}_{{\rm VaR}_{\dot{\alpha}_{\max}}(q)}y f_{\dot{\alpha}_{\max}}(y){\rm d}y\\
&=\frac{f_{\alpha}(\alpha_{\rm th})}{1-q} \int^{1}_{q}{\rm VaR}_{\dot{\alpha}_{\max}}(y)dy,
\end{align}
where the last equation is due to the standard definition of CVaR, thus yielding \eqref{ELCRmax}. Following quite a similar methodology, ELCR of $\dot{\alpha}_{\min}$ can be obtained, such as
\begin{align}
\nonumber
{\rm ELCR}_{\dot{\alpha}_{\min}}(\alpha_{\rm th};q)&=\frac{f_{\alpha}(\alpha_{\rm th})}{q}\int^{\infty}_{{\rm VaR}_{\dot{\alpha}_{\min}}(1-q)}y f_{\dot{\alpha}_{\min}}(y){\rm d}y\\
&=\frac{f_{\alpha}(\alpha_{\rm th})}{q} \int^{q}_{0}{\rm VaR}_{\dot{\alpha}_{\min}}(1-y)dy,
\end{align}
which yields \eqref{ELCRmin}.

\bibliographystyle{IEEEtran}
\bibliography{IEEEabrv,References}

\vfill

\end{document}